\documentclass[pra,aps,twocolumn,amsmath,amssymb]{revtex4}

\usepackage{graphicx}
\usepackage{dcolumn}
\usepackage{bm}
\usepackage[extension=xxx]{hyperref}

\newcommand{\beq}{\begin{equation}}
\newcommand{\eeq}{\end{equation}}
\newcommand{\beqa}{\begin{eqnarray}}
\newcommand{\eeqa}{\end{eqnarray}}

\def\beq{\begin{equation}}

\usepackage{color}

\begin{document}

\title{Fast phase-modulated optical lattice for wave packet engineering}
\date{\today}

\author{C. Cabrera-Guti\'errez, A. Fortun,  E. Michon, V. Brunaud, M. Arnal, J. Billy, D. Gu\'ery-Odelin}

\affiliation{Universit\'e de Toulouse ; UPS ; Laboratoire Collisions Agr\'egats R\'eactivit\'e, IRSAMC ; F-31062 Toulouse, France} 
\affiliation{CNRS ; UMR 5589 ; F-31062 Toulouse, France}

\begin{abstract}
We investigate experimentally a Bose Einstein condensate placed in a 1D optical lattice whose phase is modulated at a frequency large compared to all characteristic frequencies. As a result, the depth of the periodic potential is renormalized by a Bessel function which only depends on the amplitude of modulation, a prediction that we have checked quantitatively using a careful calibration scheme. This renormalization provides an interesting tool to engineer in time optical lattices. For instance, we have used it to perform simultaneously a sudden  $\pi$-phase shift (without phase residual errors) combined with a change of lattice depth, and to study the subsequent out-of-equilibrium dynamics.
\end{abstract}
\maketitle

Time-periodic driving is a powerful tool for manipulating single particle and many body systems as exemplified in the last decades in atomic physics. 

The use of rapidly oscillating electric fields, as in Paul's trap \cite{paul}, allows to circumvent the limit imposed by static Maxwell laws on the trapping of charged particles with an electric field, and is now a widely used ion trapping technique. Periodic driving has also been used successfully in cold atoms. The first Bose Einstein condensates were obtained in a trap, dubbed time-orbiting potential (TOP), in which a rapid rotating transverse magnetic field is superimposed to a static quadrupole trap in order to avoid Majorana losses \cite{TOPtrap,anderson}. Adding time-dependent magnetic fields to static traps or in combination with rf-dressing provides a wide variety of possibilities for trap geometries, and their dynamical control \cite{tiecke,hodby,Lesanovsky,garraway}. In atom chips \cite{reichel}, the rapid modulation of the current in the wires has been used to suppress the roughness of the potential due to the wire defects \cite{trebbia}. Periodic driving has also been applied to design optical potentials, using different techniques such as acousto-optic modulators and deflectors or digital-micromirror devices  \cite{friedman,onofrio,milner,zimmermann,henderson,schnelle,houston,gauthier}. 

When performed at a frequency smaller or on the order of the characteristic frequencies of the considered system, periodic driving triggers a rich out-of-equilibrium dynamics, as illustrated by the studies about quantum turbulence \cite{henn,seman,navon}.

The physics of cold atoms in optical lattices also benefits from the usefulness of time modulation.  The physics at work depends on the amplitude of modulation and on the ratio $\nu/\nu_0$ where $\nu$ is the modulation frequency and $\nu_0$ refers to the frequency associated with the first interband transition at the center of the Brillouin zone. 
For $\nu < \nu_0$, a small amplitude modulation and moderate lattice depth, the phase modulation yields the renormalization of the tunneling rate between adjacent sites \cite{rmp17}. This technique is put forward in the quantum simulation domain and enables one to generate effective Hamiltonians and engineered gauge fields \cite{PRXJean}. After the first pioneering experiments on tunneling rate renormalization \cite{ChuPRL,ArimondoPRL,Oberthaler}, a few recent experiments   have successfully exploited this technique to realize the Hofstadter model \cite{Hofstadter}, the Haldane model \cite{Haldane}, or to investigate frustrated magnetism \cite{sengstock}. 
For $\nu \sim \nu_0$ a small amplitude of phase modulation favors interband transitions \cite{raizen94,simonet} and band hybridization \cite{hy1,hy2,hy3}. In this very same range of frequencies, a large amplitude modulation generates a classically mixed phase space that was exploited for the pionnering experiments on dynamical and chaos-assisted tunneling  \cite{RaizenScience,Phillips}. 

In this article, we experimentally investigate the properties of a Bose Einstein condensate (BEC) trapped in a one dimensional optical lattice whose phase is modulated sinusoidally at a large frequency, $\nu \gg \nu_0$. We show that the depth of the periodic potential is renormalized and depends on the amplitude of modulation, allowing, for instance, to perform simultaneously an exact $\pi$ phase shift and a lattice depth change.

We have carried out our experiments on our rubidium-87 BEC machine which relies on a hybrid (magnetic and optical) trap~\cite{fortun}. We produce pure BECs of typically $10^5$ atoms in the lowest hyperfine level $F=1, m_F=-1$. The one-dimensional (1D) optical lattice is generated by the interfererence of two counter-propagating laser beams at 1064 nm (lattice spacing $d=532$ nm), superimposed to the horizontal optical guide of the hybrid trap. The relative phase of the two lattice beams is controlled through phase-locked acousto-optic modulators. We modulate the relative phase at a frequency $\nu$ and with an amplitude $\varphi_0$. In addition to the static trapping potential, $V_{\rm trap}(\vec r \,)$, associated to the hybrid trap, the atoms experience the following time-dependent potential:
\begin{equation}
V_L(x,t)= - \frac{s_0 E_L}{2} \left[1+\cos{\left(\frac{2\pi x}{d}+2\varphi_0 \sin(2\pi\nu t)\right)}\right] 
\end{equation}
where $E_L=h^2/(2md^2)$ is the lattice characteristic energy \cite{footnote1} and $s_0$ a dimensionless parameter which characterizes the lattice depth. The phase is modulated at a frequency $\nu=500$ kHz large compared to the frequency $\nu_0= 18.32$ kHz, for the lattice depth considered here ($s_0=6.40\pm 0.1$). The frequency of modulation being large compared to all frequencies of the problem, the atoms experience the effective static potential 

\begin{eqnarray}
V_{\rm eff}(\vec r \,)  & = & V_{\rm trap}(\vec r \,) + \nu \int_0^{1/\nu} V_L(x,t) dt  \\
 & = &  V_{\rm trap}(\vec r \,) - \frac{s_0E_L}{2} \left[1+ \mathcal{J}_0(2\varphi_0) \cos{\left(\frac{2\pi x}{d}\right)} \right] \nonumber
\end{eqnarray}
where $\mathcal{J}_0$ is the zeroth-order Bessel function. The effect of the high frequency modulation is therefore to renormalize the lattice depth:
\begin{equation}
s_\text{eff}=s_0\mathcal{J}_0(2\varphi_0).
\label{eqbessel}
\end{equation}
The effective depth $s_\text{eff}$ depends only on the amplitude of modulation $\varphi_0$. As intuitively expected, for large values of the amplitude of modulation, $\varphi_0$,  the lattice depth tends to zero since the oscillation washes out the periodic pattern of the potential. More generally, such renormalization allows to control the dynamics of the BEC in the lattice.

\begin{figure}[t]
\centering
\includegraphics[width=\linewidth]{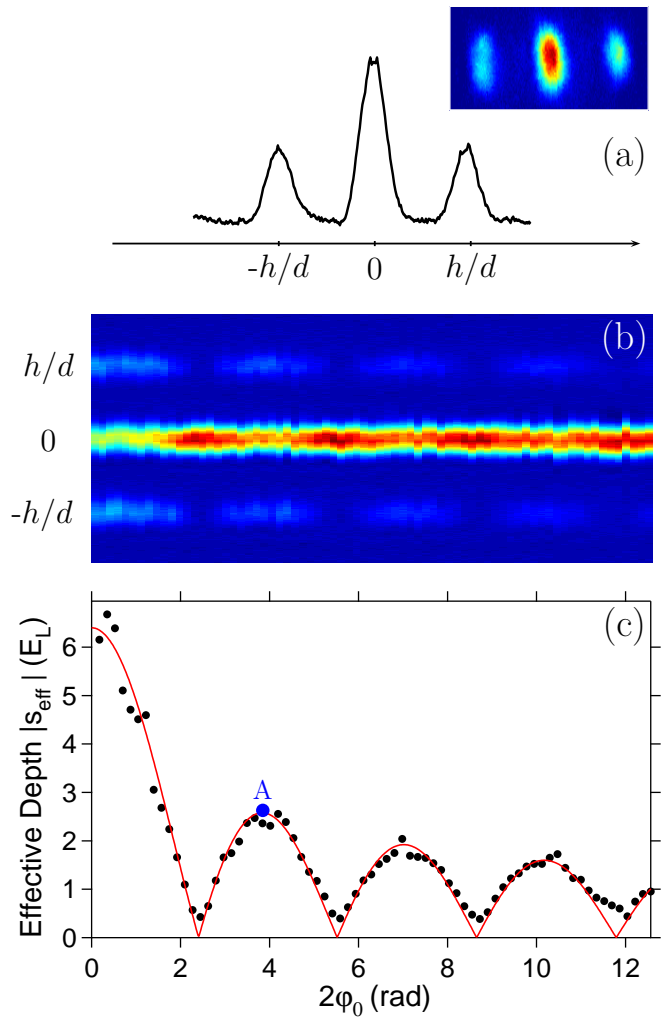}
\caption{(a) Interference pattern and its profile obtained, after a 25 ms time-of-flight, for a BEC loaded in an optical lattice of depth $6.40\pm 0.10\,E_L$. 
(b) Concatenation of interference patterns integrated along an axis perpendicular to the three peak pattern (and to the lattice direction) for BECs loaded in a phase-modulated lattice at 500 kHz. Each vertical line corresponds to a different modulation amplitude $\varphi_0$. (c) Absolute value of the effective lattice depth $\left|s_\text{eff}\right|$, extracted from the relative population of the first orders of momentum $\pm h/d$ with respect to the 0th-order, as a function of the modulation amplitude. The solid line corresponds to the theoretically expected value of $\left|s_\text{eff}\right|$ given by equation (\ref{eqbessel}) with $s_0=6.40\pm 0.1$ the experimentally measured depth of the static lattice.}
\label{fig1}
\end{figure} 

Before describing our experimental results on the renormalized optical lattice, we explain hereafter the two calibration methods we have used to determine the lattice depth.
First, we have determined the initial (static) lattice depth using a recent method based on a sudden phase shift of the lattice performed after the atoms have been loaded adiabatically in the lattice \cite{fortun,cabrera}. This shift induces the intra-site dipole motion of the atoms, which results in the oscillation of the zeroth-order population of the interference pattern obtained after a long time-of-flight (25 ms). The corresponding pattern results from the interference of the BECs trapped in each lattice well and gives access to the in-situ momentum distribution of the system. From the oscillation frequency of the population in the zeroth order (of zero momentum), we extract the lattice depth with an accuracy at the percent level, $6.40\pm 0.10\,E_L$. Interestingly, this method is immune to atom-atom interactions, the extra external confinement $V_{\rm trap}$ and remains valid if the loading of the lattice is not perfectly adiabatic \cite{cabrera}. 

The second method relies on the analysis of the interference pattern obtained after a 25 ms free time-of-flight of a BEC suddenly released from all trapping potentials and that was previously at rest in the lattice. A typical image along with its profile is given in Fig.~\ref{fig1}a.  The observed peaks are associated to the momenta $p_n= nh/d$ with $n$ an integer (positive or negative). We extract the population $\pi_n$ of the orders $n=0,\pm 1$ from the three peaks of the figure and determine the mean relative population in the first orders with respect to the population in the 0th-order: $\overline{\pi}_1=(\pi_1+\pi_{-1})/(2\pi_0)$. 

The relative populations in the different orders are commonly used to calibrate the lattice depth. However, to get a more accurate relation between the populations in the different orders of the interference pattern and the trap depth, we have revisited the standard formula \cite{cristiani} by using a systematic comparison with a numerical simulation. We calculated numerically the ground state wave function inside the lattice, and used the Fourier transform to get the correspondance between the lattice depth $s$ and the mean population in the first orders $\overline{\pi}_1$ (obtained in the absence of phase shift or for sudden phase shift with immediate release). For $\overline{\pi}_1 > 0.02$ corresponding to $s>0.5$, we find the fit function $s=\sqrt{K_0 \overline{\pi}_1}+ \sum_{n=1}^{11}K_n \overline{\pi}_1^n$ with $K_0= 14.302$, $K_1= 1.8301$, $K_2= 19.97$, $K_3=-96.533$, $K_4= 546.51$, $K_5= -995.57$, $K_6= 216.89$, $K_7=1222.1$, $K_8=486.29$, $K_9=-1490.9$, $K_{10}= -2560.1$, $K_{11}= 3486.3$, with a relative error on the estimate of the depth below 1 \% over two orders of magnitude of lattice depth ranging from $s=0.5$ to $s=50$.

To investigate experimentally the renormalization of the lattice depth, we first load adiabatically the BEC into a phase-modulated lattice for different amplitudes of modulation. After a 2 ms holding time in the phase-modulated lattice, we proceed as explained previously to obtain an absorption image with the interference pattern revealing the lattice depth effectively experienced by the atoms. We have repeated this procedure 73 times varying the amplitude of modulation $\varphi_0$ from 0 to $2\pi$. Figure~\ref{fig1}b provides a concatenation of the corresponding images where each vertical line results from an integration of the absorption image along an axis perpendicular to the three peak pattern. We observe the disappearance of the side peaks for some discrete values of the amplitude of modulation; they coincide with the zeroths of the zeroth order Bessel function $\mathcal{J}_0(x_i)=0$ for $x_i = 2.40, 5.52,8.65, 11.79, ...$. It means that for those specific values the periodic pattern is completely washed out by the modulation. 

We determine systematically the absolute value of the effective lattice depth $\left|s_\text{eff}\right|$ associated to the relative populations in the side peaks of the interference pattern obtained for different values of the phase modulation amplitude $\varphi_0$, using the fit function given above. 
The results are summarized in Fig.~\ref{fig1}c. We also plot the expected theoretical value $s_0 \left|\mathcal{J}_0(2\varphi_0)\right|$ without any adjustable parameter since the initial value is the depth of the static lattice and was determined with an independent measurement based on the sudden phase shift method as explained previously. We find a perfect agreement. Interestingly, an optical lattice with an effective depth $s_{\rm eff}$ that coincides with a local maximum of the Bessel function such as at point A in Fig.~\ref{fig1} is by definition robust against residual phase fluctuations.

\begin{figure}[h!]
\centering
\includegraphics[width=\linewidth]{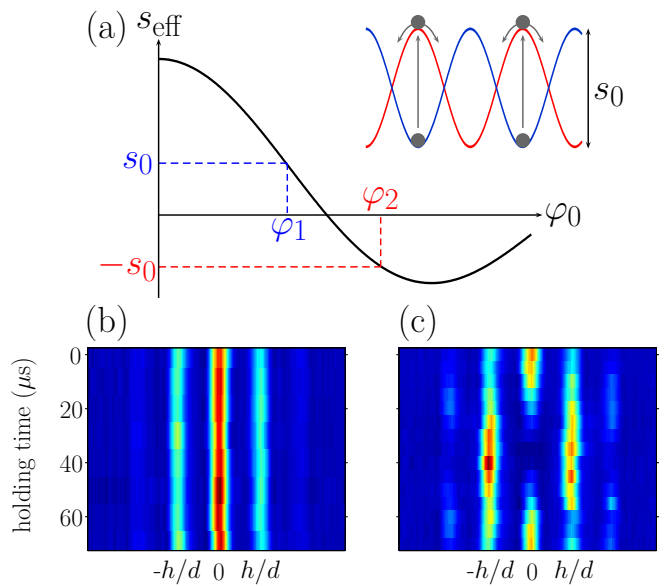}
\caption{(a) Effective depth as a function of the modulation amplitude $\varphi_0$, and sketch of the experiment in which the minima of the periodic potential are suddenly replaced by maxima as a result of a modulation amplitude change from $\varphi_1$ to $\varphi_2$. (b-c) Time-evolution of the interference pattern obtained after time-of-flight: (b) for a BEC loaded and remaining in a phase-modulated lattice with $\varphi_1=0.296 \pi$ (c) after the sudden change of the modulation amplitude from $\varphi_1$ to $\varphi_2=0.5 \pi$. Both amplitudes of modulation correspond to the same effective lattice depth, $|s_{\rm eff}|$. However, the change from  $\varphi_1$ to $\varphi_2$ triggers an out-of-equilibrium dynamics since the wave packets initially at the bottom of lattice wells are abruptly placed at the top of the periodic potential hills. Its splitting into two packets with opposite momenta is clearly seen in (c).}
\label{fig2}
\end{figure}

The sign of the Bessel function changes each time a zero is crossed. However, this property cannot be directly revealed by the interference pattern. 
In practice, it means that a sudden change in the amplitude of modulation from one side of a zero ($x_i-\varepsilon$) to the other side ($x_i+\varepsilon$ with $\varepsilon<x_{i+1}-x_i$) reverses the sign of the renormalized lattice depth, which amounts to perform a phase shift of exactly $\pi$ in addition to a change of the lattice depth. In such a change, the minima of the lattice potential for $\varphi_0=(x_i-\varepsilon)/2$ become the maxima of the modulated potential for  $\varphi_0=(x_i+\varepsilon)/2$ (see Fig.~\ref{fig2}a). 

To observe this effect, we have proceeded in the following manner. We have chosen two modulation amplitudes $\varphi_1=0.296 \pi$ ($=55^\circ$) and $\varphi_2=0.5 \pi$ ($=90^\circ$) that correspond to the same absolute value of the effective depth, around $s_{\rm eff}=s_0|\mathcal{J}_0(2\varphi_1)|=s_0|\mathcal{J}_0(2\varphi_2)|=1.95$ (see Fig.~\ref{fig2}a). First, we load the BEC into a modulated lattice with a modulation amplitude $\varphi_1$. The modulation amplitude is then suddenly switched to the value $\varphi_2$, and the system remains in this new modulated optical lattice for various amounts of time before the release, the 25 ms time-of-flight, and the absorption image. The observed interference patterns, revealing the dynamics after the change of amplitude of modulation, are shown in Fig.~\ref{fig2}c for different holding times. For the sake of comparison, we perform the same procedure without jump in the modulation amplitude: in this case the atoms simply remain at equilibrium in the minima of the lattice potential (see Fig.~\ref{fig2}b). 

The signature of the sign reversal through the change of amplitude is clearly observed in the absorption images (see Fig.~\ref{fig2}c). The sudden change in amplitude of modulation triggers an evolution in momentum space. We observe the depletion of the zeroth order of diffraction corresponding to $p=0$, and  concomitantly the increase of the populations in the first diffraction orders $p=\pm h/d$ revealing the splitting of the initial packet into two coherent packets evolving with opposite momenta.

In conclusion, we have investigated a new tool - fast phase modulation - to engineer dynamically the properties of an optical lattice. Such a tool implemented on an optical lattice of higher dimensionality would enable one to change dynamically the dimensionality or anisotropy of the optical lattice. Many other tools have been investigated in the literature to engineer quantum states in optical lattices: the Fourier synthesis of optical lattices \cite{ritt}, the control of polarization to generate spin dependent optical lattices \cite{mandel}, the control of the phase and intensity of lattice beams to generate artificial gauge fields \cite{sengstock} or  band structures with interesting topological structure  \cite{Tarruell,Hofstadter,Haldane} to name a few. Interestingly, the $\pi-$phase shift combined with lattice depth change that we have demonstrated realize the most out-of-equilibrium state reachable for a lattice of a given depth and with a wide variety of possible initial state wave functions. As it generates a splitting of the wave packets it offers the possibility to realize, at the submicron scale, an atom-atom collider. In contrast to other colliders achieved by splitting BECs \cite{wilson,walraven}, the expansion of the wave packets is here reduced which offers the possibility to study collisions from the superfluidity regime to the multiple scattering regime.

 This work was supported by Programme Investissements d'Avenir under the program ANR-11-IDEX-0002-02, reference ANR-10-LABX-0037-NEXT, and the research funding grant ANR-17-CE30-0024-01. M.~A. acknowledges support from the DGA (Direction G\'en\'erale de l'Armement).


\begin{thebibliography}{99}
\bibitem{footnote1}
The characteristic energy is given by $E_L=4E_{\rm rec}$ where $E_{\rm rec}=h^2/(2 m \lambda^2)$ is the recoil energy associated with the absorption of a single photon from one of the laser beams of the lattice.
\end{thebibliography}
\end{document}